# Enhancing interfacial magnetization with a ferroelectric

*Tricia L. Meyer,[1] Andreas Herklotz,[1] Valeria Lauter,[1] John W. Freeland,[2] John Nichols,[1] Er-Jia Guo,[1] Shinbuhm Lee,[1] T. Zac Ward,[1] Nina Balke,[1] Sergei V. Kalinin,[1] Michael R. Fitzsimmons,[1] and Ho Nyung Lee[1]\**

[1] Oak Ridge National Laboratory, Oak Ridge, TN 37831, USA
[2] Argonne National Laboratory, Argonne, IL 60439, USA
\* E-mail: hnlee@ornl.gov

Ferroelectric control of interfacial magnetism has attracted much attention. However, the coupling of these two functionalities has not been understood well at the atomic scale. The lack of scientific progress is mainly due to the limited characterization methods by which the interface's magnetic properties can be probed at an atomic level. Here, we use polarized neutron reflectometry (PNR) to probe the evolution of the magnetic moment at interfaces in ferroelectric/strongly correlated oxide [$PbZr_{0.2}Ti_{0.8}O_3$/$La_{0.8}Sr_{0.2}MnO_3$ (PZT/LSMO)] heterostructures. We find that there is always suppressed magnetization at the surface and interface of LSMO and such magnetic deterioration can be strongly improved by interfacing with a strongly polar PZT film. The magnetoelectric coupling of magnetism and ferroelectric polarization occurs within a couple of nanometers of the interface as demonstrated by the enhanced interfacial magnetization beyond the bulk value by 5% depending on the polarization of PZT. The latter value is 70% higher than the surface magnetization of a LSMO film without interfacing with a ferroelectric layer. These compelling results not only probe the presence of nanoscale magnetic suppression and its control by ferroelectrics, but also emphasize the importance of utilizing probing techniques that can distinguish between bulk and interfacial phenomena.

## I. INTRODUCTION

The doped-manganite perovskites, La$_{1-x}$Sr$_x$MnO$_3$, are perhaps one of the most well-studied classes of materials since the discovery of the colossal magnetoresistance phenomenon over six decades ago.[1] While earlier studies focused on understanding the phenomenon itself, recent studies have concentrated on nanoscale technological applications of these materials. Examples include oxide-based field effect transistors[2] and tunnel junctions[3–5] by using the intriguing metal-insulator transition (MIT) and above room temperature ferromagnetic (FM) behavior observed in several of the manganite compositions.[6] Besides the bulk properties, recent studies found many intriguing physical properties could be realized by combining these manganites with other oxides in thin film heterostructures. Among the many outstanding questions remaining, suppressed magnetization common to the surface and buried interfaces of epitaxial manganites has drawn much attention.[7–12] The deteriorated interfacial magnetic structure can reduce the polarization of the spin current due to spin-flip scattering or by being a source of poorly polarized spins. Moreover, the interface is often accompanied by an electrically insulating barrier, which influences tunneling of spin-polarized carriers essential for these applications.[13–15] Thus, understanding the origin of such dead layers and identifying ways to improve interface functionality are of great interest for fundamental research as well as technological applications.

Conventional bulk measurement techniques such as magnetometry and four point probe are often used to characterize the overall properties of thin film samples.[16–18] However, these techniques measure bulk properties rather than behavior specific to a few nanometers, *e.g.* interfacial magnetism. We have used polarized neutron reflectometry (PNR) to measure the magnetic depth profile in absolute units. The profiles are compared for cases of polarization reversal achieved by growing films in such a way that changes the interface charge density due to self-poling of the FE polarization, leading to a change in the interface magnetism.

Using PNR, many intriguing interfacial magnetic phenomena have been revealed.[11,19–23] For example, PNR has provided a better mechanistic understanding of the influence of the polar discontinuity on interface magnetism and how to eliminate it via chemically-engineered $La_{0.67}Sr_{0.33}MnO_3/SrTiO_3$ (STO) interfaces.[11,24] Additionally, PNR has been used to probe ferroelectric (FE)/magnetic interfaces.[20,25–27] The goal to change the order parameter at the boundary between different structural, magnetic or electronic ground states using a FE is indeed a highly effective approach to designing functional interfacial properties.[28–30] Since neutrons are highly penetrating and sensitive to changes in the nuclear ($n$) and magnetic ($m$) scattering length densities (SLD) of a material at the atomic scale, the chemical and magnetic evolution of buried interfaces affected by charge depletion or accumulation can be detected as changes in the $m$SLD depth profile. Considering these facts, PNR has the capability to non-destructively probe the influence of electrostatic doping upon the suppressed surface and interfacial magnetism of manganite thin films, of which the origin has been difficult to isolate (Figure 1). Previously, this behavior has been attributed to discontinuation of oxygen octahedra at the interface,[10] compositional changes,[18,22] electronic reconstruction due to polar discontinuity[11,24] or modified orbital occupancy near the film surface.[9,14] The latter explanation provided by x-ray resonant magnetic scattering and theoretical techniques is indeed compelling.

In this letter, we report the nanoscale depth-profiling of magnetic properties of $PbZr_{0.2}Ti_{0.8}O_3$ (PZT)/$La_{0.8}Sr_{0.2}MnO_3$ (LSMO) heterostructures. The FE polarization of PZT can effectively tune the interfacial charge carrier concentration of LSMO, modifying the interfacial magnetic structure. The influence on the interface magnetism is confined to a couple of nm of the FM/FE (chemical) interface. Interestingly, we find a nearly 70% enhancement of the PZT/LSMO interface magnetization compared to the uncapped LSMO film. We argue enhanced interface magnetization can be attributed to electrostatically-modified surface states of LSMO driven by charge depletion.

## II.    SAMPLE FABRICATION AND CHARACTERIZATION

Epitaxial LSMO, PZT and LaAlO$_3$ (LAO) single-crystal films were deposited on (001) SrTiO$_3$ (STO) single crystal substrates. Details of the growth are reported elsewhere.[30] The structural quality and phase purity were confirmed using x-ray diffraction (XRD) and x-ray reflectometry (Figure S1). X-ray absorption spectroscopy (XAS) spectra were collected on beamline 4-ID-C at the Advanced Photon Source of Argonne National Laboratory. Bulk magnetic measurements were performed using a Quantum Design SQUID-VSM magnetometer (Figure S2). The graded slope of the magnetization versus temperature curves and the smaller Curie temperature ($T_c$) as compared to the bulk one imply that there are multiple magnetic components to the heterostructures, whose origins can be clarified with PNR.

Specular PNR measurements were performed on the Magnetism Reflectometer at the Spallation Neutron Source of Oak Ridge National Laboratory.[31] Samples were cooled to 120 K, well below the $T_c$ of LSMO ($T_c \sim 314$ K), in an applied magnetic field of 1 T for all measurements. In PNR, the specular reflectivity ($R^\pm$) can be measured for a material, where $R^+$ and $R^-$ represent the non-spin-flip reflectivities with neutron polarization oriented parallel and antiparallel (respectively) to an external magnetic field ($H$). $R^\pm$ is measured as a function of wave vector transfer Q – the difference between the incoming and specularly reflected wave vectors. Here, we normalize the $R^\pm$ data to the asymptotic value of the Fresnel reflectivity ($R_F$).[19] The variation of the spin asymmetry [$SA = (R^+ - R^-)/(R^+ + R^-)$] with Q highlights the sensitivity to variation of the magnetization perpendicular to the FM/FE interface.[32]

## III. RESULTS AND DISCUSSION

The $R/R_F$ and $SA$ of LSMO on STO are shown in Figure 2a-b. The $n$SLD and $m$SLD depth profiles obtained from model fitting to the data are shown in Figure 2c. For the LSMO film, the $n$SLD is uniform, which suggests a nominally constant chemical composition for the total film thickness of 13.2 nm. The $m$SLD reveals three distinct regions of magnetization. Here, the substrate-film (region I) and film-vacuum interfaces (region III) exhibit suppressed magnetization values of 364 kA/m (2.4 $\mu_B$/f.u.) for 3.1 nm and 182 kA/m (1.2 $\mu_B$/f.u.) for 1.7 nm, respectively. In comparison, the largest region (region II) of the film [8.4 nm] exhibits a larger magnetization of 451 kA/m (3.0 $\mu_B$/f.u.). (Note that the PNR data fitting for the magnetization and thickness includes about 5% or less error.) The thickness for region III is consistent with earlier reports of suppressed magnetization within approximately three unit cells of the film surface.[9] Alternative fits for uniform magnetization (one magnetic region) and non-suppressed surface magnetization (two magnetic regions) clearly indicate that including suppressed surface magnetization provides the best fit (Figure S3). Given that the film is atomically flat (roughness of ~4 Å) and has an XRD rocking curve full-width-at-half-maximum value of less than 0.05°, the suppressed magnetization at the surface is likely due to factors other than structural deterioration, *e.g.* preferential formation of oxygen vacancies often reported for perovskite oxides such as STO.[33–35],

In order to test whether this surface region is sensitive to electrostatic doping, we deposited an ~8.4 nm layer of FE PZT on top of LSMO grown under identical conditions as the first sample. Using piezoresponse force microscopy, we observed that the polarization of PZT naturally points (self-poles) towards the PZT/LSMO interface. Polarization pointing towards the interface will induce hole depletion at the interface when the FM is a metal (as is the case for our LSMO film) (Figure S4). In Figure 2d−f, the PNR data for the PZT/LSMO sample are shown. In Figure 2f, the $n$SLD is similar to the uncapped LSMO film, but the magnetic profile is different. Instead of suppressed magnetization at the LSMO surface, remarkably the magnetization is enhanced to 617 kA/m (4.0 $\mu_B$/f.u.), which is larger than the magnetization for the first sample (the LSMO film without PZT).

To illustrate the confidence that we have with our fitting parameters, we have provided alternative simulations in Figure S5, which do not allow for enhanced magnetization. It is clear that neither suppressed surface magnetization nor uniform magnetization accurately explain our experimental data. In order to further support the confidence in our PNR fitting parameters, we note $M_{sat}$ obtained from magnetization versus field loops from SQUID and the integrated magnetization values determined from the different magnetic layers found by PNR in Figure S5 show excellent agreement.

The thickness of the enhanced magnetic region (region III) of 2.0(±1) nm we inferred from the PNR experiment is similar to the length scale of ~ 3-5 unit cells reported for charge screening in metallic LSMO.[4,12,21] Interestingly, the magnetization within the bulk region of the bilayer film (region II) increased from 451 kA/m (3.0 $\mu_B$/f.u) to 540 kA/m (3.6 $\mu_B$/f.u.) for the uncapped manganite film. The notable increase in magnetization for the film bulk suggests that there could be an effect of capping in addition to the interfacial magnetization enhancement from the presence of the FE. It is worth noting that capping of ultrathin films has also been shown to impact the conductivity with respect to a single layer film.[36] Capping of films can influence strain, however, reciprocal space mapping of our samples indicates that both samples are coherently strained to the substrate lattice (Figures S6). Therefore, since the strain of these samples appear the same, strain does seem a likely origin for changes of the magnetism we observed.

It is intuitive to assume that the enhanced magnetization directly at the PZT/LSMO interface is a consequence of the FE polarization, which is believed to occur on a short length scale of a few unit cells.[4,37,38] In fact, our PZT has one of the largest reported polarization values of 80 $\mu C/cm^2$, capable of inducing a change in carriers up to 0.8 $e^-$/unit cell.[39] Assuming that the magnetic interface thickness of 2 nm is proxy for the Thomas-Fermi screening length, then the PZT layer can induce a charge density in the LSMO interface of $\pm 2.5 \times 10^{21}$ cm$^{-3}$. This electrostatic doping equates to a compositional change in Sr content as high as $\Delta x = \pm 0.15$. In order to test the hypothesis that the interface magnetization is affected by electrostatic doping, we have grown a trilayer structure of two LSMO layers with a 50 nm thick PZT interlayer. This sample provides the opportunity to observe the effects of both hole accumulation and depletion at opposite interfaces. Analysis of the PNR data shown **Figure 3a−c** indicate that the magnetization of FM/FE interface (i.e., the interface formed by putting PZT on top of LSMO) is enhanced with respect to the uncapped LSMO sample (the first sample) and consistent with the enhancement observed for the second sample—the FE capped sample. Furthermore, the magnetization of the FE/FM interface (i.e., the interface formed by putting LSMO on top of PZT) is suppressed more than observed for the uncapped LSMO interface or surface. These observations are consistent with self-poling of the PZT film remaining the same as observed for the second sample, i.e., polarization pointing towards the FM/FE (bottom-most) interface. These observations agree with our expectation that hole accumulation/depletion is restricted to within a few nanometers of the interface. Due to the thickness of the PZT interlayer, we find that the PZT layer is partially relaxed, which could induce asymmetric charge-transfer screening and thereby a magnetization gradient.[25] Thus, it is likely that both strain and electrostatic effects are playing a role at the top PZT/LSMO interface.

A second test of the hypothesis that FE polarization is responsible for enhancing interface magnetism can be made by replacing the FE with a non-FE. Analysis of the PNR data for a LSMO/LAO/LSMO sample shown in Figure 3d-f found no evidence for enhanced magnetization at any interface. Therefore, the influence of PZT on the magnetic interface with LSMO is not observed with a non-FE oxide.

Nevertheless, the magnetization of the film bulk of the capped (bottom) LSMO layer (region II) in the LSMO/LAO/LSMO trilayer is increased to a magnetization of 540 kA/m (3.60 $\mu_B$/f.u.)—an increase also observed in the film bulk for LSMO capped by PZT (e.g., in the LSMO/PZT and LSMO/PZT/LSMO samples). The enhancement of the film bulk magnetization in capped LSMO is about ~0.6 $\mu_B$/f.u. compared to the same region of the uncapped LSMO film. Therefore, there are two new effects that we have identified which modify the LSMO magnetization: (1) A bulk-like enhancement extending into more than half of the film when capped (i.e. the capping layer effect) and (2) an interfacial effect occurring within a few nanometers of the FM/FE interface. As already mentioned, previous studies demonstrated that suppressed magnetization driven by polar discontinuity in manganite films can be alleviated through the incorporation of Sr-rich layers in optimally-doped $La_{0.67}Sr_{0.33}MnO_3$, resulting in improved magnetization.[11,24] In the current study, it is possible that polar discontinuity could be driving the suppressed magnetization at the STO/LSMO and LSMO/LAO interfaces, similar to other studies.[11] However, the PZT/LSMO interface exhibits a polar discontinuity comparable to the STO/LSMO substrate-film interface, which of course has suppressed magnetization. The enhanced interfacial magnetization observed when the polarization of a FE depletes holes mitigates the influence of the polar discontinuity across an interface that would otherwise suppress the interface magnetization.

In order to directly probe the effects of electrostatic doping, we have employed XAS in total electron yield mode, which is sensitive to the change in oxidation states of oxide thin films. In Figure 4, the Mn $L_3$-edge is shown. Here, the PZT/LSMO interface is compared to the surface of the LSMO film layer. The distinct peak shift by ~0.5 eV to lower energy signals a reduction in the Mn oxidation state with the addition of PZT. Further evidence for this reduction is found by the increase in peak intensity near 638 eV in the PZT/LSMO film, which is a signature of increasing $Mn^{2+}$ concentration. The reduction of the Mn ions provides clear evidence that hole depletion is taking place in the PZT/LSMO film. We attribute this behavior to hole depletion at the FM/FE interface driven by the polarization of the FE, however, oxygen vacancies can also lead to a similar XAS signature. Identifying the role of oxygen vacancies coupled with the FE and the impact of both of these parameters upon the magnetic properties is an important challenge to pursue. Regardless of whether the Mn valence is being altered by the FE polarization and/or oxygen vacancies, the salient point remains that the *interface* magnetization is affected by capping with a FE and not by a non-FE.

As mentioned previously, hole depletion via the FE's polarization is equivalent to a chemically doped composition of $x \sim 0.05$ (= $0.2 - \Delta x$), which is within the antiferromagnetic (AFM) region of the bulk manganite phase diagram. However, our results show FM is retained, and even enhanced, upon hole depletion. Enhancement of the magnetic moment per Mn atom is observed as the AF/FM phase boundary is approached from the FM side in bulk single crystals (even though $T_c$ decreases).[40] Indeed, the calculated local Mn magnetic moment for the $x = 0.05$ composition, given by $(4-x)\mu_B$, is quite close to the 4 $\mu_B$/f.u. obtained for the enhanced magnetic region in our PZT/LSMO film.[41] Nevertheless, the discrepancy between our observation that the interface remains FM for an electrostatic doping that is thought to be equivalent to chemical doping yielding an AFM ordered phase suggests that either electrostatic and chemical doping affect magnetic order differently[42] and/or the magnetic phase diagram of an interface is different than the bulk.

Interestingly, previous magnetometry studies[4,30] of PZT/LSMO heterostructures have observed that hole accumulation increases the magnetization—opposite to our findings. Others[43] have found that hole accumulation in heterostructures with similar stoichiometry led to decreased magnetization (consistent with our findings). Increased AFM interactions and concomitant loss of magnetization might be explained by a depletion of majority spin electrons, which contribute to the overall magnetic moment.[44] This explanation is well-supported by theoretical calculations of FE/manganite interfaces.[41,45,46] The question then becomes, why does hole accumulation in some studies using magnetometry (a bulk probe) indicate an overall increase in the magnetization? The answer to this question can be found when revisiting the enhanced magnetization within the bulk-like regions (region II) for films capped with PZT or LAO. The magnetization of the bulk film increased with FE or non-FE capping. The net moment is the volume-weighted moments of the film bulk (which is thick) and the interface (which is thin). Therefore, it is not surprising that magnetometry studies yield conflicting results. Thus, the value of depth-dependent techniques in discerning the role of electrostatic doping of a magnetic interface from other effects that influence the film bulk is immeasurable.

## IV. CONCLUSION

In summary, we have measured the magnetic depth profiles of PZT/LSMO ($x = 0.2$) heterostructures. We observed three effects: First, capping LSMO with a FE or non-FE oxide increases the magnetization of the LSMO film bulk. Second, the presence of the FE increases the magnetization of the FE/FM interface when the polarization of the FE points towards the interface. Third, the magnetization of the interface formed when LSMO is grown on top of the FE/FM heterostructure is suppressed due to hole accumulation induced by the FE polarization pointing away from the FE/FM interface. The thickness of the regions of the interface over which the magnetization is affected by the FE is ~2 nm. The interface effects we have observed are consistent with the influence of the electrostatic field on the hole doping of the interface affecting the magnetic behavior. These results present new and intriguing opportunities for device development, which has previously been hindered by the suppressed properties at the interface with LSMO. Moreover, this work provides direct evidence that modification of the physical properties of oxide films when interfaced can extend well beyond the interface, indicating the important role of depth profiling techniques for accurate understanding of oxide interfaces required for developing novel functional oxides.

**ACKNOWLEDGEMENTS**

This work was supported by the U.S. Department of Energy (DOE), Office of Science, Basic Energy Sciences (BES), Materials Sciences and Engineering Division [synthesis, magnetic characterization, and polarized neutron reflectometry (PNR) data analysis] and by the Laboratory Directed Research and Development Program of Oak Ridge National Laboratory (ORNL), managed by UT-Battelle, LLC, for the U. S. DOE (PNR data fitting). Use of PNR and piezoresponse force microscopy were performed as user projects at the Spallation Neutron Source and the Center for Nanophase Materials Sciences, respectively, which are sponsored at ORNL by the Scientific User Facilities Division, BES, U.S. DOE.)

Received: ((will be filled in by the editorial staff))
Revised: ((will be filled in by the editorial staff))
Published online: ((will be filled in by the editorial staff))

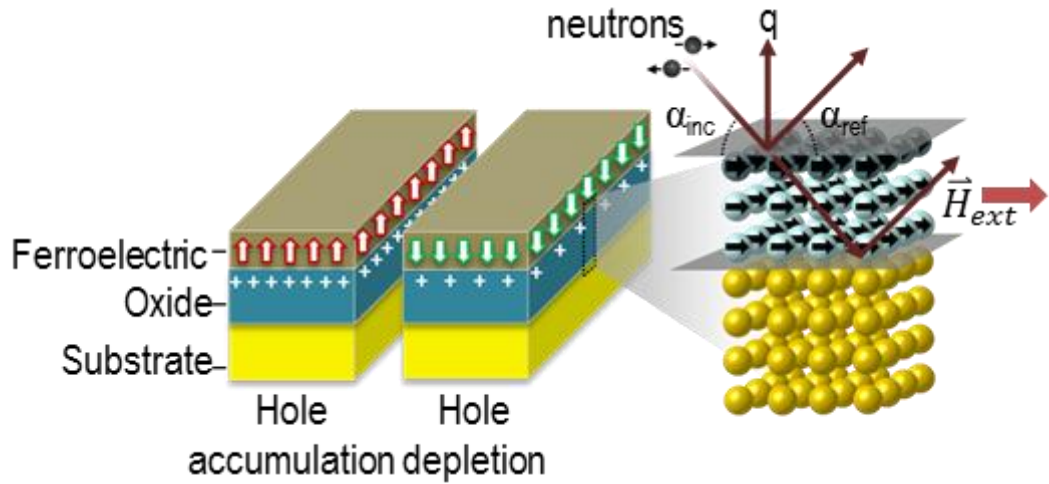

**Figure 1.** Probing buried magnetic interfaces using neutrons. Schematic of PZT-LSMO bilayer in which the polarization of PZT is switched, indicating hole accumulation and depletion. Polarized neutron reflectivity in the presence of an external magnetic field, $\vec{H}$, is used to probe interfacial modifications resulting from ferroelectric polarization. Here, $\alpha_{inc}$ and $\alpha_{ref}$ indicated the incident and reflected angle of the neutron beam and $q$ is the wave vector transfer.

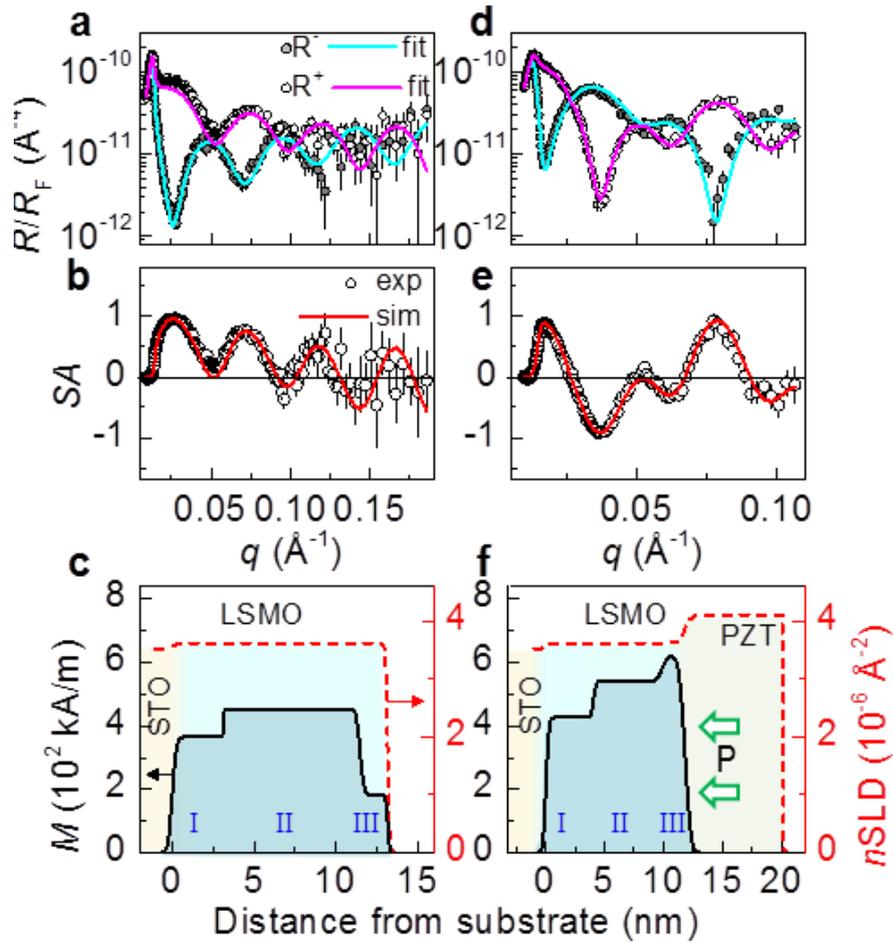

**Figure 2.** Normalized neutron reflectivity ($R/R_F$), spin asymmetry ($SA$), magnetic profile (in kA/m) and $n$SLD are shown in (a-c) for LSMO and (d-f) for PZT/LSMO layers on STO. Suppressed magnetization at the STO/LSMO interface is shown for both samples, while the LSMO/PZT interface shows enhanced magnetization. Green arrows indicate the polarization direction in the PZT layer oriented towards the LSMO film as confirmed by PFM. The gray dotted lines in (c) and (f) separate the different magnetic regions within the LSMO film layer.

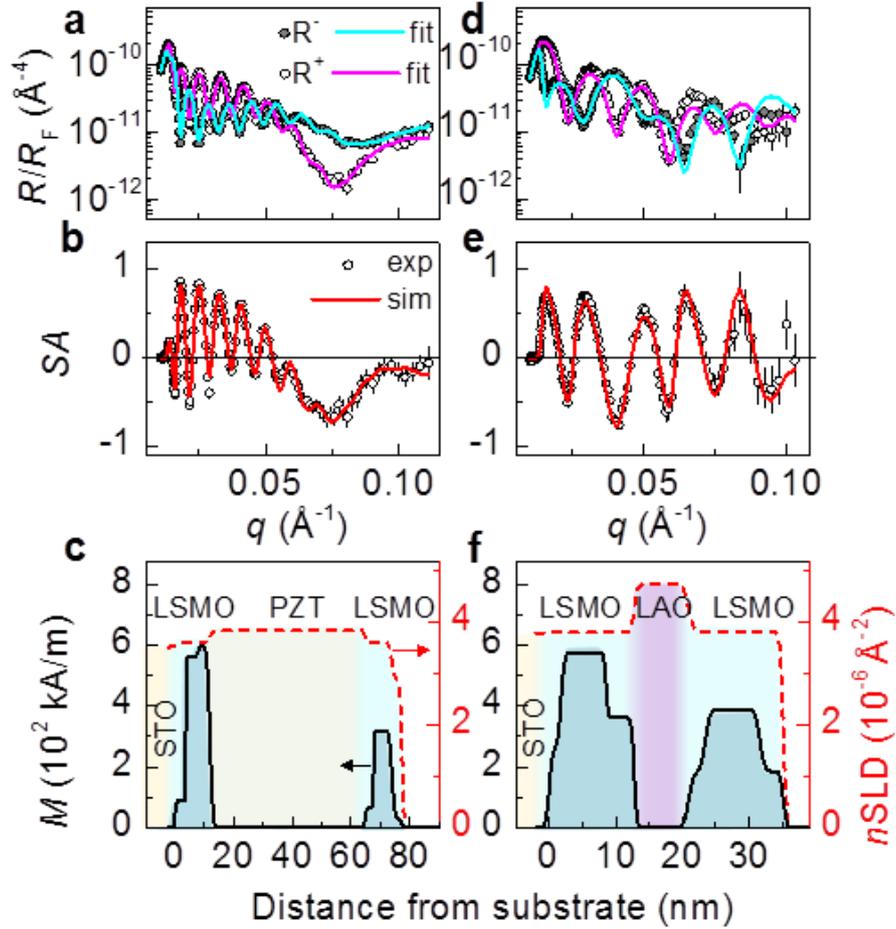

**Figure 3.** Normalized neutron reflectivity, spin asymmetry, magnetic profile and *n*SLD are shown in (a-c) for LSMO/PZT/LSMO and (d-f) for LSMO/LAO/LSMO layers on STO. Suppressed magnetization at the STO/LSMO interface is shown for both samples, whereas the LSMO/PZT/LSMO sample shows enhanced and diminished magnetization. Comparison with the LSMO/LAO/LSMO sample, which shows lower magnetization at the LSMO/LAO interface, confirms the field effect as the primary role for enhanced magnetization in LSMO in PZT/LSMO heterostructures.

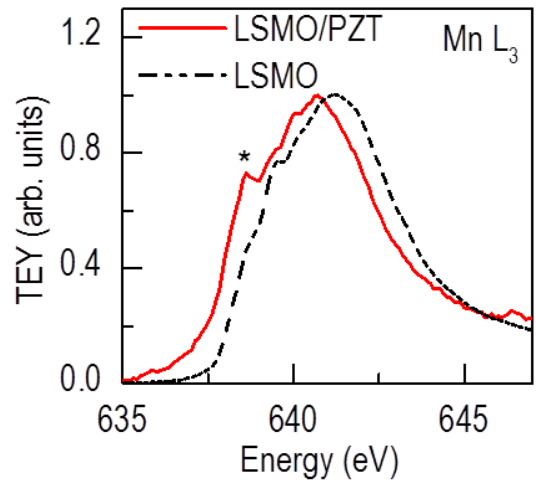

**Figure 4**. Surface sensitive total electron yield XAS spectra collected for the PZT/LSMO interface (red solid line) and the LSMO surface obtained from the LSMO/LAO/LSMO trilayer (black dashed line) sample. The peak shift to lower energy indicates reduction in the oxidation states, supported by the formation of $Mn^{2+}$ (asterisk).

**The role of ferroelectrics upon the interfacial magnetism in manganite thin films has become a subject of great interest**. Using polarized neutron reflectivity, Meyer and coworkers show that the interfacial magnetism of epitaxial manganite films can be enhanced using ferroelectric polarization, approaching large values rarely observed in thin films. Moreover, a capping layer effect was discovered, which shows that enhanced magnetization can be found away from the interface with both ferroelectric and non-ferroelectric capping layers.

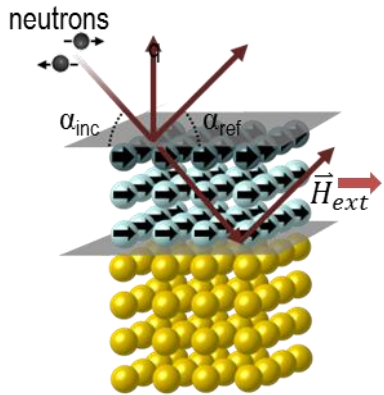